\documentclass[prl,twocolumn,superscriptaddress,nofootinbib,showpacs]{revtex4}

\usepackage{graphicx}
\usepackage{dcolumn}
\usepackage{amssymb,amsmath}
\begin{document}

\title{
Supersolid phases of a doped valence-bond 
quantum antiferromagnet: \\ 
Evidence for a coexisting superconducting order parameter}

\author{Marcin Raczkowski}
\affiliation{Marian Smoluchowski Institute of Physics, 
Jagellonian University, Reymonta 4, PL-30059 Krak\'ow, Poland}
\author{Didier Poilblanc}
\affiliation{Laboratoire de Physique Th\'eorique, CNRS and Universit\'e 
de Toulouse, F-31062 Toulouse, France}

\date{\today}

\begin{abstract}
Motivated by numerical evidence of the valence bond 
groundstate of the two-dimensional Heisenberg pyrochlore lattice, we 
argue using a $t$-$J$ model that  it evolves under doping into novel 
phases characterized by superconductivity coexisting with the underlying 
valence-bond solid order. A fermionic mean-field theory supplemented by 
exact diagonalization results provide strong arguments in favor of the 
stability of such  supersolid phases.
The resemblance with modulated superconducting patterns in high-$T_c$ 
cuprates  as well as possible relevance to frustrated noncuprate 
superconductors such as spinels and pyrochlores is discussed.
\end{abstract}

\maketitle

The so-called supersolid (SS)~\cite{Gia}, an exotic state of matter 
that breaks both the spatial lattice translation symmetry and the 
internal $U(1)$ symmetry, associated with particle number conservation, 
has attracted much attention following its possible discovery in the 
solid phase of $^4$He~\cite{He}.    
On the one hand, studies of various models with hard-core bosons 
on a frustrated triangular lattice demonstrate that bosons may condense 
forming a superfluid on top of the crystalline background~\cite{tr}. 
Moreover, there has been a convergence of agreement concerning 
nonuniform condensation of magnons in spin models with external 
magnetic field~\cite{H}, suggesting an exciting possibility of the 
formation of a spin SS in real quantum magnets. 
On the other hand, high-$T_c$ superconductivity potentially described 
by the resonating valence bond (RVB) state on the square lattice with a  
superconducting (SC) parameter of $d$-wave orbital symmetry turns out to be 
robust against spontaneous breaking of translation symmetry~\cite{ss,ss2}. 
Hence,  charge order coexisting with 
modulated SC order in Ca$_{2-x}$Na$_x$CuO$_2$Cl$_2$ revealed by  recent 
scanning tunneling microscopy~\cite{Koh07}
may well have an extrinsic origin~\cite{zn}. 

\begin{figure}[b!]
\begin{center}
\unitlength=0.01\textwidth
\begin{picture}(50,18)
\put(4,0){\includegraphics*[width=0.18\textwidth]{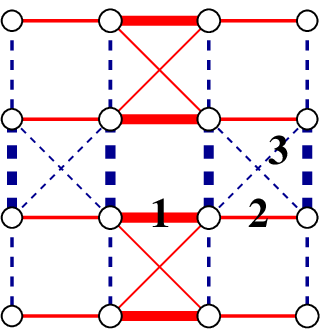}}
\put(28,0){\includegraphics*[width=0.18\textwidth]{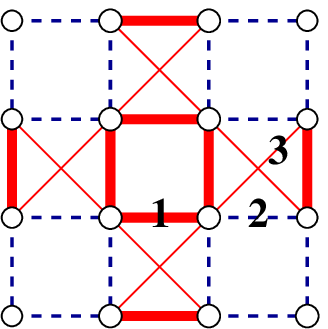}}
\put(-1.5,15.5){ {\Large (a)} }
\put(22.5,15.5){ {\Large (b)} }
\end{picture}
\end{center}
\caption {(color online)
Schematic pattern of $d_{(\pi,\pi)}$-VBSS  (a) and 
$s_{(\pi,\pi)}$-VBSS (b) phases. In the checkerboard lattice, 
diagonal bonds are on every second plaquettes.
Rotation symmetry of the pairing amplitude $\Delta_{ij}$ reduces the number 
of inequivalent bonds down to three. The line widths of these bonds labelled 
by $a$ ( $\equiv$ "$ij$" running from 1 to 3) are proportional to the magnitude 
of the corresponding spin-spin correlations (for $t'/t=-1$).
Different types of those lines (solid and dashed) correspond to opposite 
signs of $\Delta_a$. } 
\label{fig:cart}
\end{figure}

It is therefore legitimate to question the relevance of a \emph{fermionic} analog 
of the bosonic SS and its possible stabilization by geometrical magnetic frustration. 
In this respect, among of two-dimensional structures, the quantum antiferromagnet 
on the checkerboard lattice (see Fig.~\ref{fig:cart}) 
provides {\it a priori} conducive conditions to the appearance 
of a fermionic SS upon doping. 
Let us briefly summarize remarkable properties of the checkerboard
lattice: 
(i) Made out of corner-sharing plaquettes with crossed bonds, it
might be considered as a projection of a more realistic 
corner-sharing tetrahedra pyrochlore lattice known to host superconductivity, 
e.g., spinel LiTi$_2$O$_4$, as well as $\alpha$- Cd$_2$Re$_2$O$_7$ and 
$\beta$-pyrochlore KOs$_2$O$_6$~\cite{SC};  
(ii) Exact diagonalization (ED)~\cite{Fou} and other intensive studies 
of the spin-1/2 Heisenberg model  on the checkerboard lattice~\cite{pyro}
suggest that its ground state  is twofold degenerate being a product of  plaquette 
singlets. It breaks translation symmetry while preserving the $C_{4v}$ 
rotation symmetry around the center of the \emph{void} plaquette and as 
such it is a good example of a valence bond crystal (VBC)~\cite{VBC}; 
(iii) Lastly, particularly important suggestion for supersolidity 
in this case comes from  a recent discovery of hole pairing~\cite{dp04}, 
since  alone the existence of a spin gap at half-filling is 
not sufficient to guaranty SC behavior at finite doping~\cite{Leu}.

\begin{figure}[t!]
\begin{center}
\includegraphics*[width=0.4\textwidth]{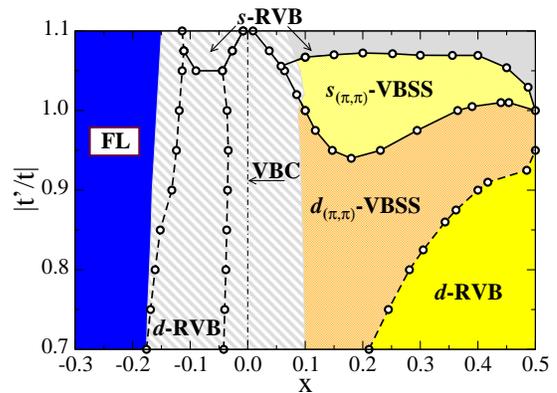}
\end{center}
\caption {(color online) RMFT phase diagram of the generalized
$t$-$J$ 
model on the checkerboard lattice with 
$t'>0$ (left) $t'<0$ (right) while $J'/J=(t'/t)^2$. 
Solid (dashed) lines show first- (second-order) phase transition, 
respectively; striped area corresponds to phase separation.  
}
\label{fig:diag}
\end{figure}

In this Letter we show that a fermionic valence bond supersolid (VBSS), 
characterized by the coexistence of the SC order parameter and solid 
component of spin singlet order, can appear by doping the VBC characterizing 
the Mott phase of some frustrated lattices like the checkerboard lattice.
Interestingly, such a scenario has close correspondence with the RVB theory 
in which the translationally invariant $d$-wave superconductor emerges out 
of the spin-liquid Mott phase.
The alternative that the VBC is destroyed upon doping 
leading to an exotic $d$+i$d$ and $d$+i$s$ 
superconductor~\cite{fc07} with broken time-reversal 
symmetry is being reconsidered~\cite{epaps}. In contrast, we find
a very rich phase diagram with various VBSS phases as summarized in
Fig.~\ref{fig:diag}. These phases differ from their bosonic 
analog not only in the way lattice symmetry is broken, exhibiting 
bond modulation rather than charge order
but, in addition, from the fact that off-diagonal and solid orders involve different 
(charge and  spin) degrees of freedom.

To address this intriguing issue, we consider a 
$t$--$J$ model on the checkerboard lattice, at doping $x=1-n$, 
$n$ being the electron density (per site):

\begin{equation}
{\cal H}= -  \sum_{\langle ij\rangle,\sigma}
    t_{ij} ({\tilde c}^{\dag}_{i\sigma}{\tilde c}^{}_{j\sigma} + h.c.)
      + \sum_{\langle ij\rangle} J_{ij} {\bf S}_i \cdot {\bf S}_j,
\label{eq:H}
\end{equation}
where the sums run over all bonds of the checkerboard lattice.
For sake of generality, we allow for different hopping amplitudes and 
antiferromagnetic superexchange couplings on the diagonal bonds, 
$t_{ij}=t'$ and $J_{ij}=J'$, and on the vertical and horizontal bonds,   
$t_{ij}=t$ and $J_{ij}=J$. 
Note that under electron-hole transformation $t'$ changes into $-t'$ 
so that hole and electron doping correspond to opposite signs of $t'$ 
(while $t>0$ can always be assumed).
Hereafter,  we assume a typical 
value $t/J=3$ and set $J'/J=(t'/t)^2$ which follows from the superexchange  
relation $J=4t^2/U$ valid in the large $U$ limit of the Hubbard model.
Next, we replace local constraints that restrict electron creation
operators ${\tilde c}^{\dag}_{i\sigma}$ to the subspace with no doubly
occupied sites by statistical Gutzwiller weights, while
decoupling in both particle-hole and particle-particle channels
yields a renormalized mean-field theory (RMFT) 
Hamiltonian~\cite{RVB},
\begin{align}
\label{eq:H_MF}
H=&- \sum_{\langle ij\rangle,\sigma} g_{ij}^tt_{ij}
                  (c^{\dagger}_{i,\sigma}c^{}_{j,\sigma}+h.c.)
                   -\mu\sum_{i,\sigma}n_{i,\sigma} \nonumber  \\
                  &-\frac{3}{4}  \sum_{\langle ij\rangle,\sigma}g_{ij}^JJ_{ij}
                  (\chi_{ji}c^{\dagger}_{i,\sigma}c^{}_{j,\sigma}
                  + h.c. -\chi_{ij}^2) \nonumber  \\
                  &-\frac{3}{4}  \sum_{\langle ij\rangle,\sigma}g_{ij}^JJ_{ij}
                  (\Delta_{ji}c^{\dagger}_{i,\sigma}c^\dagger_{j,-\sigma}    
                  + h.c. -|\Delta_{ij}|^2),
\end{align}
with the Bogoliubov-de Gennes self-consistency conditions for the bond-
$\chi_{ji}=\langle c^\dagger_{j,\sigma}c^{}_{i,\sigma}\rangle$ 
(assumed here to be real) and pair-order
$\Delta_{ji}=\langle c^{}_{j,-\sigma}c^{}_{i,\sigma}\rangle
            =\langle c^{}_{i,-\sigma}c^{}_{j,\sigma}\rangle$
parameters in the unprojected state that we solve on a 48$\times$48 
cluster at low temperature $\beta J=200$. Note that since we restrict 
our study to strong frustration ($|t'|\simeq t$), we do not consider here  
long-range antiferromagnetic order, known to coexist with superconductivity 
at low doping on the square lattice~\cite{AF}.
Finally, to allow for small nonuniform charge modulations (if any), 
the Gutzwiller weights have been expressed in terms of local 
doped hole densities 
$ x_{i}=1-\sum_{\sigma}\langle c^{\dagger}_{i\sigma}c^{}_{i\sigma}\rangle$
as $g_{ij}^t=\sqrt{z_i z_j}$ and
$g_{ij}^J=(2-z_i)(2-z_j)$ with $z_i=2x_{i}/(1+x_{i})$.

We mostly concentrate here on the $t'<0$ case  
with emergent supersolidity. Guided by the earlier studies at 
half-filling~\cite{Fou,pyro},  in addition to usual $d$- and $s$-RVB 
superconductivity 
we also allow for coexisting VBC order in the mean-field equations.
The $d$-wave symmetry \cite{RVB}  is a natural choice compatible with 
the required  $C_{4v}$ rotation symmetry of $|\Delta_{ij}|$ 
around the centers of void plaquettes 
(bonds labeled by 1 in Fig.~\ref{fig:cart}).
It yields at half-filling a strong 
insulating VBC with decoupled void plaquette 
singlets (i.e. $\chi_a=\Delta_a=0$ on the $a=2,3$ bonds)
which provides the lowest magnetic energy $E_J/J=-3/4$
in good agreement with numerics \cite{Fou}. 
Upon finite doping $x$, an additional $\Delta_{3}$ component arises along the 
diagonals as depicted in Fig.~\ref{fig:cart}(a). Unit cell doubling together 
with $d$-wave symmetry implies  a ($\pi,\pi$) phase modulation of 
$\Delta_{3}$ so that this phase is referred to as $d_{(\pi,\pi)}$-VBSS.  
In fact, away from half-filling,
another SS state, labeled by $s_{(\pi,\pi)}$-VBSS and shown in 
Fig.~\ref{fig:cart}(b), is competing with the $d_{(\pi,\pi)}$-VBSS. 
In this phase, all neighboring \emph{void} plaquettes have opposite signs of 
the pairing parameters and $s$-wave rotation
symmetry on every void plaquette.

\begin{figure}[t!]
\begin{center}
\includegraphics*[width=0.38\textwidth]{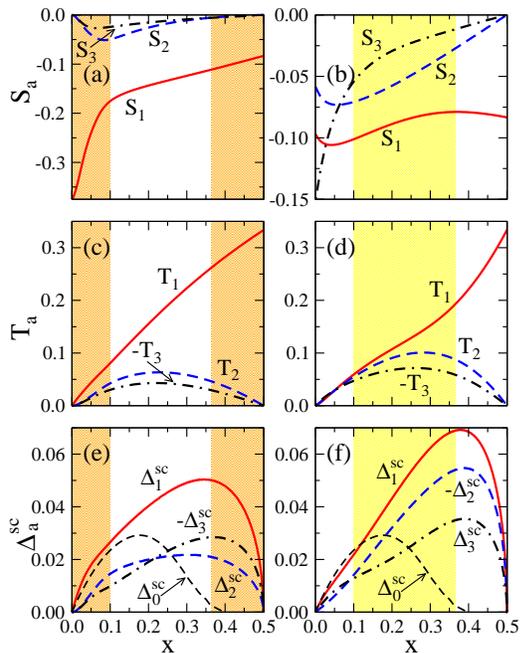}
\end{center}
\caption {(color online)
(a,b) Spin-spin correlations $S_{a}$ and (c,d) bond charge hoppings $T_{a}$ 
on the three inequivalent bonds of the  $d_{(\pi,\pi)}$-VBSS (left) and 
$s_{(\pi,\pi)}$-VBSS (right) phase for $t'/t=-1$. 
Panels (e,f) show the SC order parameters $\Delta_{a}^{\rm SC}$ 
found  to be enhanced with respect to $\Delta^{\rm SC}_0=\Delta^{\rm
  SC}_1=\Delta^{\rm SC}_2$  
on a simple square lattice ($t'=0$), see also Ref.~\cite{fc07}.
Shaded areas denote the region(s) of relative stability of each phase.
}
\label{fig:ord2}
\end{figure} 

\begin{figure}[b!]
\begin{center}
\includegraphics*[width=0.38\textwidth]{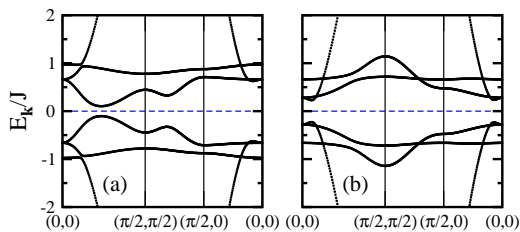}
\end{center}
\caption {(color online)
Quasiparticle band structure found for $t'/t=-1$ at $x=1/4$: 
(a) $d_{(\pi,\pi)}$-VBSS and (b)  $s_{(\pi,\pi)}$-VBSS. 
} 
\label{fig:bs}
\end{figure}

The spontaneous lattice symmetry breaking
of the discovered  VBSS phases is seen in Fig.~\ref{fig:ord2}(a-d) 
showing bond charge hopping
$T_{a}^{}=2g_{a}^{t}{\chi_{a}^{}}$ and spin-spin correlations
$S_{a}=-\frac{3}{2}g_{a}^{J}(\chi_{a}^2+|\Delta_{a}^{}|^2)$ 
(corresponding to the two terms of Eq.~(\ref{eq:H}), respectively)  on 
the three inequivalent bonds defined in Fig.~\ref{fig:cart}. 
The $s_{(\pi,\pi)}$-VBSS is characterized by a weaker 
modulation of $S_{a}$ as compared to its $d$-wave counterpart. 
Even though the $d$-wave RVB order is known to 
yield a better kinetic energy than the $s$-wave RVB one~\cite{RVB}, 
the $s_{(\pi,\pi)}$-VBSS has better total kinetic energy $E_t$~\cite{epaps} 
due to the absence of phase modulation in the pairing $\Delta_3$. 
Next, in Fig.~\ref{fig:ord2}(e,f), we present the SC order 
parameter components $\Delta_{a}^{\rm SC}$ which are related in RMFT to pairing 
amplitudes by $\Delta_{a}^{\rm SC}=g_{a}^{t}\Delta_{a}^{}$ and hence vanish
as expected at half-filling \cite{RVB}. Remarkably, all of them show a 
non-monotonic dome shape as a function of doping. 
Although the Gutzwiller RMFT is a zero temperature scheme, we expect 
that SC properties will be boosted to higher temperatures compared to 
the $d$-RVB phase with $t'=0$.  Note that VBC and SC orders 
can give rise to separate critical $T_c$. Finally, the low-energy  fermionic
quasiparticle band structures of the SS phases are shown in
Fig.~\ref{fig:bs}. Interestingly,  in the $d_{(\pi,\pi)}$-VBSS, 
the $\Delta_3$ $s$-wave component on the diagonal bonds suppresses the usual 
zero-energy Dirac cone of the $d$-wave superconductor along its $(0,0)$-$(\pi/2,\pi/2)$
nodal direction. 
Upon further doping, a gapless spectrum is recovered  in the $d$-RVB phase
with $\Delta_3=0$ due to melting of the valence bond order.
In contrast, the energy gap of the $s_{(\pi,\pi)}$-VBSS is less 
modulated as expected for isotropic pairing.

\begin{figure}[t!]
\begin{center}
\includegraphics*[width=0.38\textwidth]{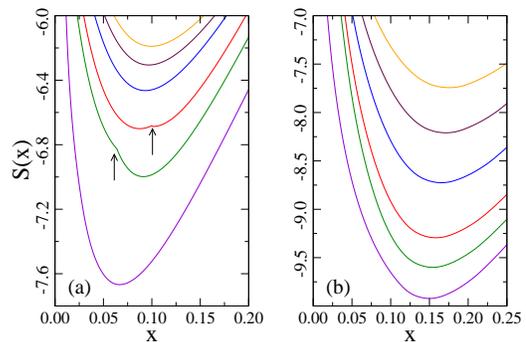}
\end{center}
\caption {(color online)
Maxwell construction (see text) for $t'<0$ (a) and $t'>0$ (b). 
From top to bottom: $|t'/t|$=0.7, 0.8, 0.9, 1.0, 1.05, and 1.1.
The arrows indicate small kinks due to the first order transition expected 
if PS ({\it cf.} Fig.~\ref{fig:diag}) is not considered. 
} 
\label{fig:ps}
\end{figure}

It is known that doping a Mott insulator can lead to phase separation (PS), 
i.e., a macroscopic segregation into two phases, commonly
the undoped phase and a phase with a finite doping~\cite{Emery}.
Hence, we have performed the standard Maxwell construction 
and calculated the slope of the ground state energy density 
$\mathcal{S}(x)=[E(x) - E(0)]/x$ as a function of doping $x$.
Technically, PS is signaled by a negative curvature
of $E(x)$ when $x\to 0$ (i.e. a negative compressibility)
and a minimum of $\mathcal{S}(x)$ at a given doping $x_c$ 
({\it cf.} Fig.~\ref{fig:ps}).
We expect that below $x_c$ the system phase
separates into a two-component mixture of the undoped ($x$=0) VBC
and a doped phase (see the phase diagram) with $x=x_c$. Note that
other phase separated mixtures which could in principle appear in the 
vicinity of some first-order transitions within the $0<x<x_c$ region 
(seen e.g. as kinks in Fig.~5) are here prevented by the PS between the 
above-mentioned "extremal'' phases.

We now briefly discuss the case $t'>0$. First, under doping, the emergent 
$d_{(\pi,\pi)}$-VBSS is now much more fragile w.r.t. the melting of the VBC
order. Secondly, a smaller 
range of stability is found for the $d$-RVB phase compared to the previous $t'<0$ case 
consistently with the strong electron-hole asymmetry of pairing 
found in numerics~\cite{dp04,note_conventions}.
On top, as deduced from the Maxwell construction shown in 
Fig.~\ref{fig:ps}(b), these narrow VBSS and $d$-RVB regions appear to be
unstable towards PS between a hole-free VBC and a homogeneous metallic Fermi 
liquid (FL) with $x=x_c\simeq 0.15$ at $t'=t$.

Our findings are summarized in a global RMFT phase diagram 
in the ($x$, $|t'/t|$) plane reported in Fig.~\ref{fig:diag}. It emphasizes 
the crucial role of the frustration in stabilizing the VBSS states for 
$t'<0$. Indeed, increasing frustration strength $|t'/t|$ shifts the second-order 
phase transition to the nearby $d$-wave RVB phase towards 
higher doping level. A region of  $s_{(\pi,\pi)}$-VBSS is squeezed into a narrow range 
$0.94\lesssim |t'/t|\lesssim 1.07$ limited below (above) by a first-order  
phase transition line to the $d_{(\pi,\pi)}$-VBSS ($s$-RVB) phase.
In fact, for maximum frustration, both $d_{(\pi,\pi)}$- and 
$s_{(\pi,\pi)}$-VBSS states extend up to quarter-filling $x=1/2$ where 
they merge to a VBC similar to the one at half-filling 
(albeit with no pairing, i.e., $\Delta_a=0$). 
Note that in the whole PS region, the role of the so far neglected 
long-range Coulomb repulsion will become important and lead to 
microscopic (possibly stripe-like) PS instead of true macroscopic PS.
Also, although quantum fluctuations can play an important role, 
here we can nevertheless provide simple additional arguments 
supporting our mean-field phase diagram. First of all, the 
large spin gap (estimated by ED of a 32-site
cluster to be around 0.69 $J$ \cite{Fou}) is  
expected to survive at least in the vicinity of 
half-filling. Next, we have found that the two-hole binding 
energy estimated by ED of the same cluster is substantial 
even in the J/t $\rightarrow$ 0 limit where it scales as $-$4.4396 J 
(with $|t'|$=t), confirming strong pairing.
Finally,  we have checked that below $|t'/t|=0.4$, the VBSS melts towards 
phase-separated $d$-RVB state in agreement with Variational Monte Carlo 
simulations for $t'=0$ \cite{Dima}.

Experimentally, examples  where spatial frustration and superconductivity coexist 
are KOs$_2$O$_6$ and LiTi$_2$O$_4$. Although electron correlations are usually 
associated to nodal pairing, recent experiments show the evidence for fully gapped 
superconductivity in both compounds~\cite{Os}, so conventional 
phonon-mediated pairing has been advocated.  
However, our results indicate that, 
on a strongly frustrated lattice, electron correlations can 
also lead to gapful pairing and a modulated structure.  
We also note that, despite interesting similarities, the $C_{4v}$-symmetric 
VBSS states differ from the unidirectional $C_{2v}$ SS found on the 
square lattice of Ca$_{2-x}$Na$_x$CuO$_2$Cl$_2$~\cite{Koh07} 
by the absence of spatial modulation in the local hole density.      

In summary, we have shown that the interplay between electron 
correlations and geometrical frustration can stabilize
novel states of matter exhibiting  microscopic coexistence of
superconductivity and spin dimer-crystalline order. 
In contrast to the inhomogeneous superconductor of the 
frustrated Shastry-Sutherland lattice which is disconnected from the 
half-filled dimer ground state~\cite{Liu},  {\it fermionic supersolids}  emerge 
naturally upon doping a parent Mott insulator with spontaneous spin dimer order.
We have identified on the checkerboard lattice two different 
competing supersolids with either $s$-wave or $d$-wave pairing symmetry 
w.r.t. the centers of the void plaquettes. 
We believe that our findings open a new route 
to search for new exotic frustrated superconductors.

\begin{acknowledgments}
We thank Fu-Chun Zhang for valuable discussions at an early stage of this 
work. M.R. acknowledges support from the Foundation for Polish Science 
(FNP) and from Polish Ministry of Science and Education under Project 
No. N202 068 32/1481. D.P. acknowledges support from the French Research 
Council (ANR).

\end{acknowledgments}

\newpage

\section*{ 
Supplementary data to the article:
Supersolid phases of a doped valence-bond
quantum antiferromagnet: Evidence for coexisting superconducting order
}

In our Letter, we have provided strong arguments in favor of
the formation of a fermionic valence bond 
supersolid (VBSS) on the checkerboard lattice. It is a new alternative 
scenario to the one in which the valence bond crystal (VBC) known 
to set in at half-filling \cite{Fou,pyro} is destroyed upon doping while 
frustration produces an exotic $d$+i$d$ and $d$+i$s$ superconductivity  
with broken time-reversal symmetry~\cite{fc07}. Hence, we have applied the
renormalized mean-field theory (RMFT) to study 
the stability of various VBSS structures on equal footing with the 
$d$+i$d$ and $d$+i$s$ superconductors that preserve the translation 
symmetry of the lattice.   
The results for two representative cases: half-fillling $x=0$ and $x=1/4$ 
doping (the latter chosen to be outside the expected phase separated region) 
are summarized in 
Table~\ref{tab}. They clearly show that the half-filled Mott insulating 
ground state corresponds to the VBC with $d$-wave (spin) pairing 
amplitude and has substantially better energy as compared to the next
low-energy $d$+i$d$ order. Under doping it evolves into  
$d_{(\pi,\pi)}$-VBSS (in fact, thermodynamically unstable towards phase 
separation 
below a small critical doping). Further doping results in a first-order phase 
transition towards a new $s$-wave superconducting ground state which 
is found to coexist with (spin dimer) crystalline order forming thus 
the $s_{(\pi,\pi)}$-VBSS. 
Note also that around $x=1/4$ a simple homogeneous 
$s_{(\pi,\pi)}$-RVB phase\footnote[1]{Similar "plaquette'' SC state appears 
on the Shastry-Sutherland lattice, see B.-J. Yang, Y. B. Kim, J. Yu, and 
K. Park, \prb {\bf 77}, 104507 (2008).}
with the same pairing symmetry as the $s_{(\pi,\pi)}$-VBSS 
(but with $|\Delta_1|=|\Delta_2|$ and $\Delta_3=0$) also 
becomes very competitive.

\begin{table}[b!]
\caption {
Site-normalized RMFT energy $E$ and kinetic energy $E_t$ of various 
wave functions considered in our Letter obtained for $t'/t=-1$. 
Pure $s$-RVB and $d$+i$s$ phases converge at half-filling 
to a homogeneous Fermi liquid; $d$+i$d$ phase is unstable at $x=1/4$ 
towards $d$-RVB.
} 
\begin{ruledtabular}
\begin{tabular}{cccccc}
\multicolumn{2}{c}{}  &\multicolumn{1}{c}{$x=0$} &\multicolumn{1}{c}{} 
                      & \multicolumn{2}{c}{$x=1/4$}\cr
\multicolumn{1}{c}{phase}  &\multicolumn{1}{c}{} & 
\multicolumn{1}{c}{$E/J$}  & \multicolumn{1}{c}{}&
\multicolumn{1}{c}{$E/J$}  &\multicolumn{1}{c}{$E_t/J$} \cr
\colrule
$s$-RVB           && $-$       && $-$2.0848 & $-$1.7928  \cr
$d$+i$s$          && $-$       && $-$2.0868 & $-$1.7859  \cr
$s_{(\pi,\pi)}$-RVB && $-$0.5944 && $-$2.1044 & $-$1.7964  \cr
$s_{(\pi,\pi)}$-VBSS/VBC  && $-$0.6134 && $-$2.1062 & $-$1.7945  \cr
$d$-RVB           && $-$0.6884 && $-$2.0776 & $-$1.7752  \cr
$d$+i$d$          && $-$0.6886 && $-$       & $-$        \cr
$d_{(\pi,\pi)}$-VBSS/VBC  && $-$0.7500 && $-$2.0980 & $-$1.7729  \cr 
\end{tabular}
\end{ruledtabular}
\label{tab}
\end{table}


\begin{thebibliography}{00}

\bibitem{Gia} T. Giamarchi, C. R\"uegg, and O. Tchernyshyov, 
              Nature Physics {\bf 4}, 198 (2008).

\bibitem{He}  E. Kim and M. H. W. Chan, 
              Nature (London) {\bf 427}, 225 (2004);
              Science {\bf 305}, 1941 (2004).

\bibitem{tr} S. Wessel and M. Troyer, \prl {\bf 95}, 127205 (2005);
             D. Heidarian and K. Damle, {\it ibid.} {\bf 95}, 127206 (2005);
             R. G. Melko, A. Paramekanti, A. A. Burkov, A. Vishwanath, 
             D. N. Sheng, and L. Balents, 
             {\it ibid.} {\bf 95}, 127207 (2005).
            

\bibitem{H}  K.-K. Ng and T. K. Lee, \prl {\bf 97}, 127204 (2006);
          P. Sengupta and C. D. Batista, {\it ibid.} {\bf 98}, 227201 (2007); 
             N. Laflorencie and F. Mila, {\it ibid.} {\bf 99}, 027202 (2007).

\bibitem{ss}  M. Raczkowski, M. Capello, D. Poilblanc, R. Fr\'esard, 
              and A. M. Ole\'s, 
              \prb {\bf 76}, 140505(R) (2007);  M. Capello, M. Raczkowski, 
               and D. Poilblanc, 
              {\it ibid.\/} {\bf 77}, 224502 (2008). 

\bibitem{ss2} C.-P. Chou, N. Fukushima, and T. K. Lee, \prb {\bf 78}, 134530 (2008);
              K.-Y. Yang, W.-Q. Chen, T. M. Rice, M. Sigrist, and 
              F. C. Zhang, arXiv:0807.3789.


\bibitem{Koh07} Y. Kohsaka, 
                C. Taylor, K. Fujita, A. Schmidt, C. Lupien, 
                T. Hanaguri, M. Azuma, M. Takano, H. Eisaki, H. Takagi, 
                S. Uchida, and J. C. Davis, 
                Science {\bf 315}, 1380 (2007).

\bibitem{zn}  R. K. Kaul, R. G. Melko, M. A. Metlitski, and S. Sachdev,
              \prl {\bf 101}, 187206 (2008).


\bibitem{SC} H. Aoki, J. Phys.: Condens. Matter {\bf 16}, V1 (2004).


\bibitem{Fou} J.-B. Fouet, M. Mambrini, P. Sindzingre, and C. Lhuillier, 
             \prb {\bf 67}, 054411 (2003). 

\bibitem{pyro} E. Berg, E. Altman, and A. Auerbach,
               \prl {\bf 90}, 147204 (2003); 
               O. Tchernyshyov, O. A. Starykh, R. Moessner, and A. G. Abanov, 
               \prb {\bf 68}, 144422 (2003); 
               J.-S. Bernier, C.-H Chung, Y. B. Kim, and S. Sachdev,
               {\it ibid.} {\bf 69}, 214427 (2004); 
              for related studies in the Hubbard model, see: 
               T. Yoshioka, A. Koga, and N. Kawakami, 
               \prb {\bf 78}, 165113 (2008).


\bibitem{VBC} N. Read and S. Sachdev, 
              \prl {\bf 62}, 1694 (1989). 

\bibitem {dp04} D. Poilblanc,
             \prl {\bf 93}, 197204 (2004).

\bibitem{Leu} See, e.g., ED studies on the frustrated Shastry-Sutherland lattice: 
              P. W. Leung  and Y. F. Cheng,
              \prb {\bf 69}, 180403(R) (2004).

\bibitem{fc07} H.-X. Huang, Y.-Q. Li, J.-Y. Gan, Y. Chen, and F. C. Zhang,
               \prb {\bf 75}, 184523 (2007). 


\bibitem{epaps} See EPAPS Document No. for supplementary data. 
                For more information on EPAPS, see 
                http://www.aip.org/pubservs/epaps.html.

\bibitem{RVB}    F. C. Zhang, C. Gros, T. M. Rice, and H. Shiba,
                 Supercond. Sci. Technol. {\bf 1}, 36 (1988).

\bibitem{AF}  T. Giamarchi and C. Lhuillier, \prb {\bf 43}, 12943 (1991);
              A. Himeda and M. Ogata,  {\it ibid.\/} {\bf 60}, R9935 (1999); 
              L. Spanu, M. Lugas, F. Becca, and S. Sorella, 
              {\it ibid.\/} {\bf 77}, 024510 (2008).

	
\bibitem{Emery} V. J. Emery, S. A. Kivelson, and H. Q. Lin,
                \prl  {\bf 64}, 475 (1990).

\bibitem{note_conventions} In Ref.~\cite{dp04} 
{\it hole} convention is used so that hoppings are opposite to the 
{\it electron} hoppings defined here.


\bibitem{Dima} D.~A. Ivanov, 
               \prb {\bf 70}, 104503 (2004). 

\bibitem{Os} C. P. Sun \emph{et al.}, 
              \prb {\bf 70}, 054519 (2004);
             Y. Kasahara \emph{et al.}, 
             \prl {\bf 96}, 247004 (2006).

\bibitem{Liu} J. Liu, N. Trivedi, Y. Lee, B. N. Harmon, 
               and J. Schmalian, \prl {\bf 99}, 227003 (2007).




\end{thebibliography}
\end{document}